# Amorphous Intergranular Films Enable the Creation of Bulk Nanocrystalline Cu-Zr with Full Density

Olivia K. Donaldson[1] and Timothy J. Rupert[1*]

[1]Department of Material Science and Engineering, University of California Irvine, CA 92697 USA
[*]Email address: trupert@uci.edu

Nanocrystalline metal alloys show great potential as structural materials, but are often only available in small volumes such as thin films or powders. However, recent research has suggested that dopant segregation and grain boundary structural transitions between states known as complexions can dramatically alter grain size stability and potentially enable activated sintering. In this study, we explore strategic consolidation routes for mechanically alloyed Cu-4 at.% Zr powders to capture the effects of amorphous complexion formation on the densification of bulk nanostructured metals. We observed an increase in density of the consolidated samples which coincides with the formation of amorphous intergranular films. At the same time, the grain size is reasonably stable after exposure to these temperatures. As a result, we are able to produce a bulk nano-grained metal with a grain size of 57 nm and a density of 99.8%, which shows an impressive balance of small grain size and high density using simple consolidation techniques.

Nanocrystalline materials, commonly defined as having grains less than 100 nm, show significant potential as structural materials due to their high strength,[1, 2] wear resistance,[3-5] and fatigue lifetime[6] that is achieved through a dramatic reduction in grain size.[7-12] Although nanocrystalline materials exhibit many desirable properties, their large volume fraction of grain boundaries result in thermal instability.[13] The energetic penalty associated with the grain boundaries can even drive low temperature grain growth in some situations.[14-16] When grains coarsen to above the nanocrystalline regime, the targeted improvements to strength and other properties have been lost. Undesirable coarsening can occur in service, but also during the simple act of making the materials. Many processing routes, including the sintering needed to consolidate nanostructured powders,[7, 17, 18] expose materials to combinations of heat and pressure that can result in grain growth. Since nanocrystalline materials are often created in limited volumes,[19] the ability to consolidate to create bulk pieces is essential for the intergration of these materials into real-world applications.

To counteract the thermal instability issues, stabilization of the nanocrystalline microstructure can be achieved through alloying. Alloying can act to restrict boundary migration either through thermodynamic or kinetic considerations,[16] yielding a stable grain structure. Grain boundaries can be modified thermodynamically, as explained in the work of Chookajorn et al.,[20] when the segregation of a dopant to the boundaries corresponds to a reduction of the grain boundary energy. While thermodynamics drive the segregation, the segregant can also provide kinetic stabilization through Zener pinning. As the dopant decorates the boundaries, it restricts the migration of the boundary and limits grain growth. Grain boundary segregation that stabilizes nanocrystalline grain structures[16, 21-23] has been observed in a multitude of systems such as W-Ti,[24, 25] Ni-P,[26, 27] Fe-Mg,[28] Cu-Ta,[29] and Au-Pt.[30]

In addition to a simple reduction in energy, doped boundaries can undergo structural transformations. These complexions can be categorized in a number of ways, such as by geometry, structure, composition, or thickness.[31] These complexion structures can range from a boundary with simple submonolayer segregation to discrete phase-like layers with nanoscale equilibrium thicknesses.[32] If one chooses to look at structural order, nanoscale layer complexions can be either ordered or disordered. Disordered or amorphous intergranular films (AIFs) are particularly interesting, as they have been shown to enhance the ductility of nanocrystalline metals.[33] AIFs could theoretically form in pure systems[31] but all observations to date have relied on grain boundary doping to induce premelting below the bulk solidus temperature.[34] These AIFs form because they have a lower energy than the traditional grain boundary (i.e., they are thermodynamically more stable), leading to improved grain size stability in some cases.[35] The work of Schuler and Rupert[32] demonstrated design rules for selecting alloys which could sustain AIFs. These authors found that the critical ingredients for AIF formation were (1) segregating dopants and (2) a negative enthalpy of mixing.[32] The Cu-Zr system is one of the model systems for AIF formation, with early studies providing evidence of strong Zr segregation to the boundaries and the formation of thick AIFs at high temperatures.[32, 33, 36-38]

Separate powder metallurgy studies have implicated AIFs in the activated sintering phenomenon, where enhanced diffusion is achieved.[39, 40] Luo and Shi[41] analyzed the theromdynamics of AIFs and found that AIF formation aligned with the onset of sintering in well-known systems such as ZnO doped with $Bi_2O_3$ [42], Ni-W[40, 43], and $Ce_{0.9}Gd_{0.1}O_{1.95}$ doped with cobalt-oxide.[44, 45] Sintering is fundamentally a trade-off between two events: (1) densification of the powder and (2) coarsening of the grains inside the powder particles. A given process tends to optimize either one (full dense, but coarse-grained material[17, 46-48]) or the other (a porous nanocrystalline piece[49, 50]) variable. A path to a bulk nanostructured metals requires fast consolidation yet a stable grain structure. Although the observations of activated sintering and grain size stabilization have to date come from separate studies, we hypothesize that they can both be active at the same time and allow for the creation of a fully dense and fine grained nanostructured metal. In this study, the sintering of Cu-4 at.% Zr powder, an alloy system that has been shown to form AIFs, was characterized to determine the ideal conditions for the production of a dense nanocrystalline structure. A specific focus was placed on temperatures in the range of and above where AIF formation has been reported. Our main goal is to determine whether AIF formation and existence during sintering simultaneously stabilizes grain size and improves densification.

Ball milling was used to create nanocrystalline Cu-Zr powders. Energy dispersive spectroscopy (EDS) inside of a scanning electron microscope (SEM) was implemented and an alloy composition of 3.62 at.% Zr was measured. Since EDS typically has an error of ± 1 at.%, we round up and refer to this sample as Cu- 4 at.% Zr from now on. The mechanical alloying process resulted in a distribution of particle sizes, as shown in Figures 1a and b. The particle sizes ranged from 12 to 248 μm, with an average particle size of 53 μm. The particle size distribution is log-normal, with 93% of the particles smaller than 100 μm. Having some variety of particle sizes is generally beneficial for the creation of a dense sintered sample as the smaller particles fill in the spaces between the larger particles during cold compaction.[51] X-ray diffraction (XRD) of the powder confirmed that the powder particles had nanocrystalline grains, with an average grain size (*d*) of 35 nm in the as-milled condition. In addition, the alloyed powders were found to contain a small amount of ZrC, approximately 4 vol.%, as a result of reactions between the Zr powders and the processing agent that was added to prevent agglomeration and cold welding. Such unintentional precipitation is commonly observed in ball milled alloys.[19]

Powders were either cold pressed at 25 MPa by hand or at 100 MPa using a hydraulic press. XRD after this processing step showed that no grain growth occured in either case. Most of the green bodies were then sintered for 1 h at a pressure of 50 MPa, labeled as Condition A in Table 1 and all subsequent figures. The sample densities were determined by cross-sectioning and imaging with the SEM to determine the density of the samples following ASTM standards, E 1245-03,[52] with the results for the Condition A samples shown in Figure 2a. In this study, we only focus on relatively high temperatures of 500 °C and above, as dense pieces were sought. For sintering temperatures of 500-700 °C, densities of ~97% were achieved with little variation as a function of temperature. These densities are generally below the value of at least 99% commonly used in the literature to describe a compact as "fully dense".[53] However, when the annealing temperature goes higher to >800 °C, there is a sharp increase in the density of the sintered samples to 99.4-99.9%. Luo and Shi predicted that the grain boundary disordering to form AIFs could occur at temperatures roughly in the range of 60-85% of the melting temperature,[41] with the high end of this value being 785 °C for Cu-4 at.% Zr (marked in Figure 2a by the maroon line).[34] An experimental report of the temperature range for AIF formation can be obtained from the work of Khalajhedayati and Rupert.[36] These authors studied grain boundary structure in Cu-3 at.% Zr alloys that were heat treated at various temperatures and then rapidly quenched to freeze in any high temperature interfacial structures. They found that ordered doped grain boundaries were observed after annealing at temperatures up to and including 750 °C but then observed AIFs when the samples were annealed at 850 °C, providing an expected AIF formation temperature between these two values (marked as a grey region).[36] The temperatures given by both the theoretical prediction and the direct characterization of boundary structure coincide with the jump in density of our sintered samples in Figure 2a. As a result, we can conclude that AIF formation does in fact allow for activated sintering and a much improved densification process. We note that applied pressure during hot pressing could in theory affect the critical temperature for AIF formation. However, the applied pressure is relatively low at only 50 MPa and the rapid rise in sample density closely aligns with prior measurements and predictions of the critical temperature, implying that such an effect is not operating here. A summary of all sintering parameters tested and the resultant densities is presented in Table 1. Additionally, more complicated thermal treatments such as the usage of an annealing step before hot pressing, which other authors have postulated will remove air pockets and improve final density[51], were examined and showed no strong effect.

Consolidation to full density is only a notable achievement if the nanocrystalline microstructure was retained, so we next studied the effect of consolidation temperature on grain size as measured by XRD (Figure 3a). Minimal grain growth was found up to ~900 °C, after which the grain structure begins to coarsen. We do note that the grain growth observed at 900-950 °C is still less than the grain growth observed in pure nanocrystalline Cu.[54-56] Most importantly, there is only a small amount of grain growth found in the range of 800-850 °C, where AIFs first form and activated sintering was found. The AIF formation predictions from the works of Luo and Shi[41] and Khalajhedayati et al.[36] are again shown in this figure. The formation of AIFs within the Cu - 4 at.% Zr alloy was confirmed by annealing a powder from the same milling batch at 850 °C for 1 h, followed by quenching, with a high resolution TEM image of an AIF shown in Figure 3d. The XRD grain sizes were confirmed with transmission electron microscopy (TEM) bright field imaging, with cumulative distribution functions of grain size for three samples and a representative TEM image shown in Figures 3b and 3c, respectively. These values are also provided for comparison in Table 1. The lack of rampant grain growth was at least partially the result of AIF formation and not solely resulting from the existence of secondary phases, as the amount of ZrC present in the sample is not enough to limit grain growth through Zener pinning [36] and X-ray and electron diffraction confirmed that no intermetallics phases were present. We do note that the carbide particles do aid the thermal stability of the alloys, with the ZrC particles and AIFs likely being complementary features for this alloy. A range of processing conditions including different cold pressing pressures, different hot pressing times, and the use of pressure-free annealing were explored while keeping the hot pressing temperature at 900 °C, which are shown in Figure 3a. While these variables did not greatly affect the density of the samples (Table 1), there were noticable changes to the final grain size. Generally, it appears that cold pressing with a higher pressure does slightly help by reducing the final grain size, while other variations tend to have a negative affect and result in coarser final materials. In addition, most powders were sieved before consolidation, following the suggestions of MTI Corporation, but a few non-sieved powder sets were also hot pressed. Table 1 shows that using non-sieved powders actually gives a slightly smaller grain size and comparable density in the final compact when all other variables are kept constant.

The results of Figures 2 and 3 show that our original goals, densification without severe grain growth, were achieved with the help of AIFs in nanocrystalline Cu-Zr. For comparison, we compile a collection of data from the literature where both density and grain size were directly reported in Figure 4, to compare with the results of our study. In previous studies where metal grains remained nanocrystalline after consolidation, most of the reported densities were much lower in the range of 40 to 90%.[49, 50, 57, 58] On the other hand, fully dense or nearly fully dense metals usually had grains in the ultra-fine grained regime with $d$ above 100 nm. It is important to note that the optimization of density and grain size in metallic systems is generally behind work in ceramics, where fully dense specimens with grain sizes below 20 nm have been made with spark plasma sintering (SPS) and environmentally controlled pressure assisted sintering (EC-PAS). Notable work in this area includes studies from Muche et al.[59], Sokol et al.[60], and Ryou et al.[61]. For example, Park and Schuh[17] used nano-phase separation in W-Cr and W-Ti-Cr to produce ultra-fine grained structures with grain sizes ranging from 97 to 846 nm, with densification to reasonably high values of 90 to 98.6% through nano-phase separation. This nano-phase separation was characterized by the rapid diffusion of a solid phase and the decoration of an inter-particle neck.[17] Other systematic studies of the trade-off between density and grain size under various sintering conditions have been carried out in Y-rich alloys[49], $BaTiO_3$,[50] Cu-Cr,[58] Ni-Cu-Zn,[57] $TiO_2$, and $ZrO_2$.[62] In general, the compiled data shows that densification usually comes at the expense of grain size, with fully dense metals (commonly defined as relative density values ≥99%[53] and marked with a red line in Figure 4) typically having grain sizes in the ultra-fine grained or coarse-grained regimes. $BaTiO_3$ and W-Cr in Figure 4 show this behavior with grain sizes ranging from 100 to 850 nm.[17, 50] Similar results have been observed in pure Cu, 90% dense[63], which can be further enhanced with the addition of refractory metals. Density was found to increase through the addition of Ru, giving densities from 97.2 to 99.89%[64], while Ta additions also limit grain growth and result in full densification.[65] While many sintering processes result in coarse-grained samples, work on the Cu-Ta system by Darling et al.[66] and Hammond et al.[67] have reported on nanocrystalline materials with grain sizes of 50 and 84 nm, respectively. While these studies have produced nanograined sintered samples described by the authors as "fully dense," the exact density values are not reported in the papers. To allow for a comparison, this data is plotted in Figure 4 at 99% density (shown in blue on Figure 4). We reiterate that these density values could be different from the real values but only place this data here due to our lack of additional knowledge about the samples. While the Cu-Ta alloys reach a comparable combination of density and grain size, equal-channel angular extrusion (ECAE) was used for consolidation. This process is more difficult to scale to larger sample sizes and generally requires more complex tooling. The Cu-4 at.% Zr samples studied in this work show an impressive balance of high density and small grain size reported to date, as shown in the zoomed view presented in Figure 4b, with the added benefit that only simple consolidation equipment was used. The optimal combination of a 57 nm grain size and a 99.8% relative density was achieved after hot pressing our powders for 1 h at 850 °C under 50 MPa of pressure. This sample's hardness was determined using nanoindentation with an Agilent G200 system using 50 mN maximum loads, giving an average hardness of 4.5 GPa which is roughly two times higher than nano-grained pure Cu[68]. The combination of a very fine nanocrystalline grain structure and achievement of full density open the way for bulk nanostructured metals with high strengths to be used for real engineering applications.

In summary, mechanically alloyed Cu-4 at.% Zr samples that were sintered into bulk pieces were analyzed to determine how Zr segregation and AIF formation affects the competition between densification and grain growth during consolidation. An impressive combination of 57 nm grain size and 99.8% density was achieved, much improved over past literature reports. Most importantly, a rapid increase in compact density was found as AIFs began to form in the microstructure, directly implicating these features in the activated sintering process. Ultimately, this study indicates that it is possible to stabilize dense nanocrystalline bulk metal structures at elevated temperatures through the utilization of AIFs.

**Experimental Section**

Nanocrystalline Cu-4 at.% Zr powders were created through mechanical alloying using a SPEX 8000M Mixer Mill. The Cu (99% pure) and Zr (99.5% pure) powders were combined in a hardened steel jar with steel media and mechanically alloyed for 10 h in an inert atmosphere (99.99% Ar). The alloying employed a 10:1 ball to powder ratio and 1 wt.% stearic acid was added as a process control agent. Following milling, the powders were cold pressed to form a green body using a graphite die in either (1) a 15 metric ton laboratory press from MTI Corporation for a total of 10 minutes at 25 MPa or (2) a hydraulic press from Enterpac for a total of 10 minutes at 100 MPa. Next, hot press sintering was performed using a vacuum heated pressing furnace, OTF-1200X-VHP4 from MTI Corporation, under a vacuum pressure of -0.1 Pa and with a heating rate of 10 °C/min. All sintered materials are therefore the result of a simple consolidation treatment. Conditions for the hot pressure sintering were varied, with details listed in Table 1, to examine the effects of annealing temperature and pressing time on densification.

After consolidation, the samples were allowed to cool to room temperature, which typically took 2-4 hours. Initial characterization was performed with a Rigaku SmartLab X-ray Diffractometer operated at 40 kV and 44 mA, using a Cu cathode. Rietveld analysis was used to determine the average grain size of each sample and XRD measurements of grain size were confirmed with TEM imaging. TEM samples were created with the focused ion beam (FIB) lift-out method using a FEI Quanta 3D FEG Dual Beam SEM/FIB. Bright field images of the consolidated samples were taken using JEOL JEM-2100F and JEOL 2800 TEMs and at least 100 grains were measured for each sample. The density of the consolidated samples were determined using porosity calculations following ASTM standard E1245-03. Hardness was determined by running load control nanoindentation experiments on an Agilent G200 Nanoindenter with a Berkovich diamond tip, a maximum load of 50 mN, and a 20% to unloading step.

**Acknowledgements**

This research was supported by the U.S. Army Research Office under Grant W911NF-16-1-0369. XRD and TEM work was performed at the UC Irvine Materials Research Institute (IMRI). SEM and FIB work was performed at the UC Irvine Materials Research Institute (IMRI) using instrumentation funded in part by the National Science Foundation Center for Chemistry at the Space-Time Limit (CHE-0802913). The authors would like to thank Dr. Alexander D. Dupuy and Professor Julie M. Schoenung of the University of California, Irvine for their help with the hydraulic pressing. In addition, the authors would like to thank Jenna L. Wardini for her help with the analysis of the carbide particles in the mechanically alloyed powder.

| Sample Type | Consolidation Conditions | XRD Grain Size (nm) [TEM Grain Size (nm)] | Relative Density (%) [Standard Error] |
|---|---|---|---|
| *Cold Pressed 25 MPa 10 minutes (sieved powder)* | | | |
| A | 500 °C for 1 hr at 50 MPa | 43 [37] | 96.9 [0.42] |
| A | 600 °C for 1 hr at 50 MPa | 44 | 97.2 [0.54] |
| A | 700 °C for 1 hr at 50 MPa | 48 | 96.6 [0.45] |
| A | 800 °C for 1 hr at 50 MPa | 56 | 99.4 [0.08] |
| A | 850 °C for 1 hr at 50 MPa | 57 [65] | 99.8 [0.06] |
| A | 900 °C for 1 hr at 50 MPa | 82 | 99.9 [0.03] |
| A | 950 °C for 1 hr at 50 MPa | 107 | 99.9 [0.05] |
| C | 900 °C for 10 hrs at 50 MPa | 111 | 99.7 [0.24] |
|  | 1 hour annealed at 900 °C and pressed at 500 °C for 10 hours at 50 MPa | 81 | 99.9 [0] |
| D | 1 hour annealed at 500 °C then pressed at 900 °C for 10 hrs at 50 MPa | 104 [85] | 99.8 [0.11] |
| E | 1 hour annealed at 500 °C then pressed at 900 °C for 5 hrs at 50 MPa | 95 | 99.98 [0.01] |
| *Cold Pressed 25 MPa 10 minutes (non-sieved powder)* | | | |
|  | 600 °C for 1 hr at 50 MPa | 39 | 97.9 [0.83] |
|  | 900 °C for 1 hr at 50 MPa | 62 | 99.6 [0.18] |
|  | 600 °C for 10 hrs at 50 MPa | 41 | 99.7 [0.12] |
|  | 900 °C for 10 hrs at 50 MPa | 85 | 99.7 [0.06] |
| *Hydraulic Press 100 MPa 10 minutes* | | | |
| B | 900 °C for 1hr at 50MPa | 72 | 99.9 [0] |
| G | 1 hour annealed at 500 °C and pressed at 900 °C for 10 hours at 10 MPa | 105 | 99.9 [0.05] |
| F | 1 hour annealed at 500 °C and pressed at 900 °C for 10 hours at 50 MPa | 115 | 99.98 [0.01] |

**Table 1.** Processing conditions for sintered samples with corresponding average grain size and relative density measurements.

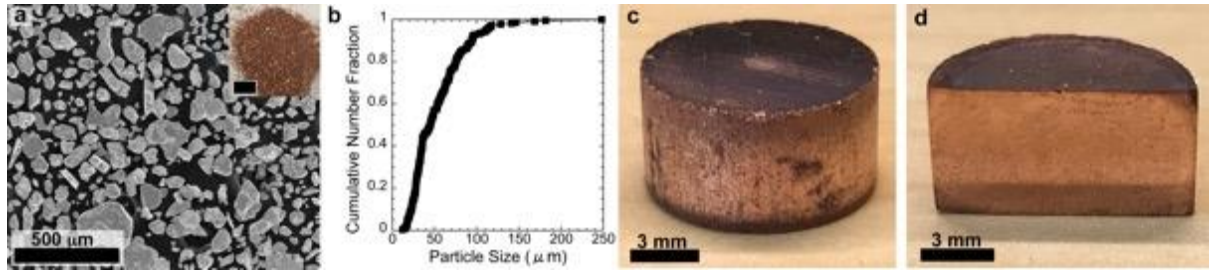

**Figure 1.** (a) SEM image of the mechanically alloyed Cu-4 at. % Zr powder, with an optical image of a powder pile shown in the inset (scale bar is 5 mm in the inset), and (b) cumulative number fraction graph distribution function of the particle size of the as-milled powder. Optical images of the (c) complete sintered sample and (d) polished cross-section of the consolidated sample.

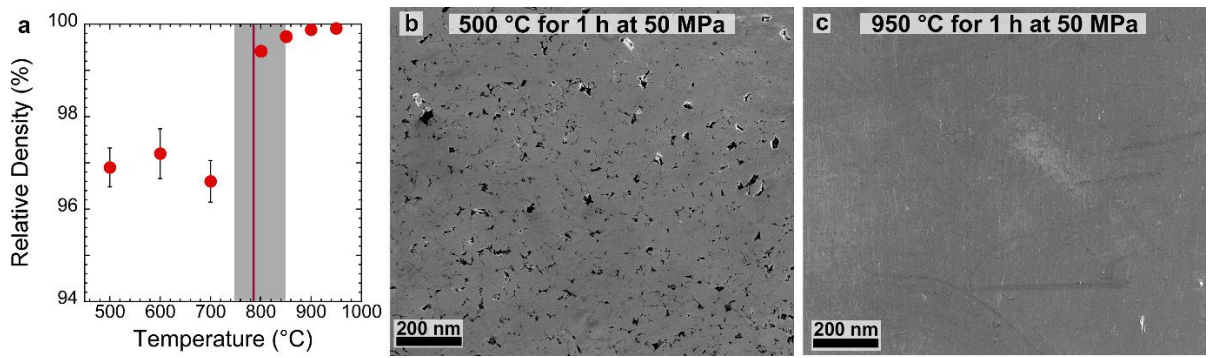

**Figure 2.** (a) Relative density as a function of pressing temperature for 1 h at an applied pressure of 50 MPa. A jump in density was observed at 850 °C, which aligns with experimental identification of the AIF formation temperature range in Ref [36], marked by the grey rectangle, and the theoretical prediction (85% of the melting temperature) from Ref[41], marked by the maroon line. Standard error has been included for all of the density measurements. Error bars are present but difficult to see above 800 °C because they are small relative to the data point size. SEM images of sintered samples were used to measure the density of the consolidated samples, with examples from (b) 500 °C for 1 h at 50 MPa and (c) 950 °C for 1 h at 50 MPa conditions shown here.

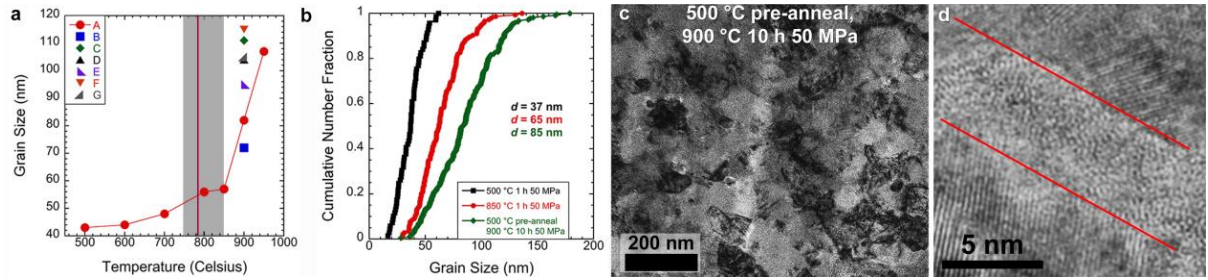

**Figure 3.** (a) Grain size as a function of pressing temperature graph, showing limited grain growth up to ~900 °C. (b) Cumulative distribution functions for the grain sizes measured by TEM for three of the sintered specimens. (c) Bright field TEM image of the 500 °C pre-annealed, 900 °C for 10 h at 50 MPa grain structure. (d) An AIF, outlined by the red lines, is shown within a Cu-4 at.% Zr powder sample that was annealed at 850 °C for 1 h and then subsequently quenched.

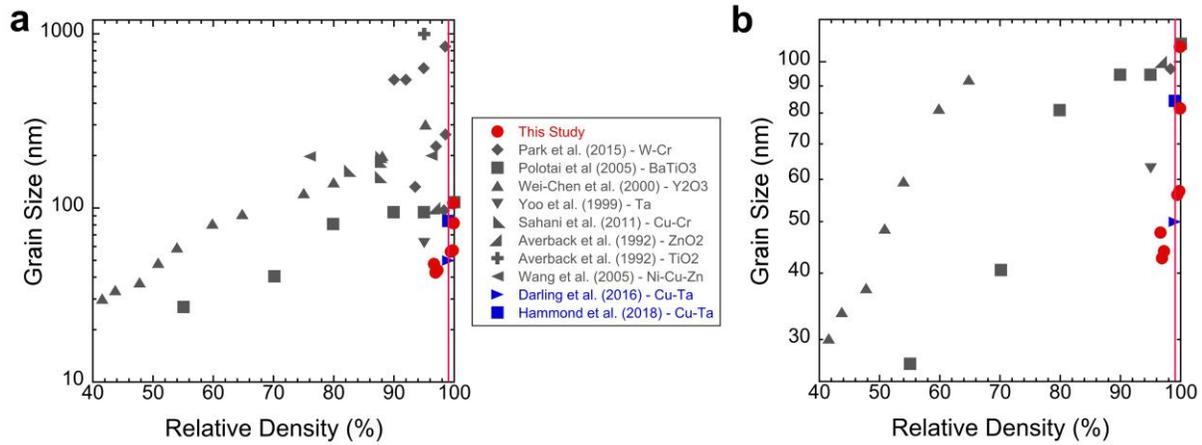

**Figure 4.** (a) Grain size as a function of relative density for both the sintering samples from this study and values from the literature[17, 49, 50, 57, 58, 62, 66, 67, 69], with a zoomed view of the best data shown in (b). The nanocrystalline Cu-Zr samples created in this study achieve a better combination of high density and small grain size than any other study. The red line in both figures denotes a density of 99%, which is commonly used as a definition of "fully dense".